\documentclass[aps,reprint,prb,showpacs]{revtex4-2}
\usepackage[utf8]{inputenc}

\usepackage{subfigure}
\usepackage{array}
\usepackage{amsthm}
\usepackage{amsmath}
\usepackage{amssymb}
\usepackage{slashed}
\usepackage{bm}
\usepackage{graphicx}
\usepackage{wrapfig}
\usepackage{physics}
\usepackage{mathtools}
\usepackage{subfiles}
\usepackage{color, colortbl}
\usepackage{hyperref}
\usepackage{marginnote}
\usepackage{caption}
\usepackage{subcaption}
\usepackage{xr}
\usepackage{float}

\newcommand{\inb}[3]{{\left#1 #2 \right#3}}
\newcommand{\reff}[1]{{Fig.\ref{fig:#1}}}
\newcommand{\refe}[1]{{Eq.\ref{eq:#1}}}

\date{}

\begin{document}
	\title{Harris-Luck criterion in the plateau transition of the Integer Quantum Hall Effect}
	\author{H. Topchyan $^{1}$,  W.Nuding$^{2}$, A. Kl\"umper$^{2}$  and A. Sedrakyan$^{1}$}
	\affiliation{$^1$ A.Alikhanyan National Science Laboratory, Br. Alikhanian 2, Yerevan 0036, Armenia \\	
		$^2$Wuppertal University, Gau\ss stra\ss e 20, 42119 Wuppertal, Germany}
	
	\begin{abstract}		

The Harris criterion imposes a constraint on the critical behavior of a system
upon introduction of new disorder, based on its dimension $d$ and localization
length exponent $\nu$.  It states that the new disorder can be relevant only
if $d \nu < 2$. We analyze the applicability of the Harris criterion to the  GKNS
network disorder formulated in the paper [I. A. Gruzberg, A. Kl\"umper,
W. Nuding, and A. Sedrakyan, Phys. Rev. B 95, 125414 (2017)] and show that the
fluctuations of the geometry are relevant despite $d \nu> 2$, implying that
Harris criterion should be modified.  We have observed that the fluctuations
of the critical point in different quenched configurations of disordered
network blocks is of order $L^0$, i.e.~it does not depend on block size $L$ in contrast to the expectation based on the Harris criterion that they should decrease as $L^{-d/2}$ according to the central limit theorem.
Since $L^0 > (x-x_c)$ is always satisfied near the critical point, the mentioned network disorder is relevant and the critical indices of the system can be changed.
We have also shown that the GKNS disordered network is fundamentally different from Voronoi-Delaunay and dynamically triangulated random lattices:
the probability of higher connectivity in the GKNS network decreases in a power law as opposed to an exponential,
indicating that we are dealing with a ``scale free" network, such as the Internet, protein-protein interactions, etc.

	\end{abstract}
	
	\date{{\today}}
	
	\pacs{
		71.30.$+$h;% Metal-insulator transitions and other electronic transitions
		71.23.An;  % Theories and models; localized states
		72.15.Rn   % Localization effects (Anderson or weak localization)
	}
	\maketitle

\section{Introduction}
The Harris criterion \cite{Harris-1974, Chayes-1986} is a condition for the
stability of the phase transition when a new disorder field is added
to the problem. If this condition is fulfilled, the fluctuations of the fields
near the critical point caused by the new disorder are not strong enough to change
the critical indices. This means that the additional disorder is
irrelevant. The principle is as follows: consider a system
with control parameter $x$ and a corresponding critical point $x_c$. If the model
has additional random parameters, where $N$ is the number of such parameters
(which is related to the size of the system), then the position of the
critical point will also change, $x_c \rightarrow x_c + \Delta x_c$. According
to the central limit theorem, the order of fluctuations of the critical point
is expected to be $\Delta x_c \propto N^{-1/2}$.

If we study a lattice with disordered fields, the number of random parameters is given by $N \propto L^d$, where $L$ is the linear size of the considered system and $d$ is its dimension. We can divide the system into subsystems of correlation length $L=\xi$, and assign to each of the subsystems its own value of the critical point $x_{c,i}=x_c + \Delta x_{c,i}$. The fluctuation of the critical point $\Delta x_c = \inb<{\Delta x_{c,i^2}}>^{1/2}$ (which can be interpreted as the width of the positions of the critical values resulting from different configurations of the subsystems) will be of order $\xi^{-d/2}$. Assuming that the system is in some state with parameter value $x$, then expressing the correlation length on the ``distance" from the critical point as $\xi \propto (x - x_c)^{-\nu}$ leads to the expression
\begin{equation}
    \Delta x_c \propto (x - x_c)^{d \nu / 2} \quad.
\end{equation}
Starting from a state near criticality,
% with $|x-x_c| \gg \Delta x_c$,
it is obvious that if $d \nu / 2 > 1$, the condition $\Delta x_c \ll |x-x_c|$
holds continuously when $x\rightarrow x_c$, and it does not hold in the
opposite case. This argument holds for uncorrelated, homogeneous, and
isotropic disorders. If $d \nu / 2 < 1$, we will have $\Delta x_c \gg
|x-x_c|$, which means divergent behavior between different subsystems. This
can lead to a change in the overall critical behavior (including the critical
indices). In that case we will say that the critical behavior of the system
is unstable against random defects, in other words, additional disorder
is relevant. In the opposite case, the critical behavior is stable, so impurities cannot change the critical indices. This condition is known as the Harris criterion.

This analysis contradicts the results presented in \cite{Sedrakyan-2017,
  Sedrakyan-2019,Sedrakyan-2024}. In these papers, a modification of the two dimensional Chalker-Coddington (CC) network model \cite{Chalker-1988} has been formulated based on geometrical fluctuations of the network, which randomly changes its shape and introduces curvature to the originally flat CC model. Since in the CC model we have the critical index $\nu \simeq 2.56$ \cite{Slevin-2009, Sedrakyan-2011,Beenakker-2011,Beenakker-2-2011,Obuse-2012,Slevin-2012,Sedrakyan-2015}, the Harris criterion for the relevance of additional disorder is not fulfilled, which predicts that the index should not be changed, contrary to the statements of the papers \cite{Sedrakyan-2017,Sedrakyan-2019,Sedrakyan-2024}. This raises the question of the validity of the Harris criterion in the case of  two dimensional geometric disorder, namely, disorder of the network. The answer to this question {concerning 2d spaces} is the main goal of the current work. 

Violations of the Harris criterion have been discussed in several papers \cite{Halperin-1983,Prudnikov-1999, Luck-1993,Okabe-2000,Schliecker-2002}. In the presence of disorder correlations, the criterion is no longer directly applicable, but it can be generalized accordingly. In the case of long-range, algebraically decaying correlations, the condition of relevance depends on the dimension of the disorders as well as on the power of decay of the correlations \cite{Halperin-1983,Prudnikov-1999}. For another disorder not covered by the random-bond paradigm, namely the coordination number aperiodicity found in quasicrystals and other aperiodic structures, Luck \cite{Luck-1993} formulated a modification of the relevance criterion by introducing a new exponent called ``wandering". He argued that if the new disorder is initiated by a modulated geometric structure, which can be viewed as a deformation of a regular lattice, then the Harris criterion can change. The argument is based on the hypothesis that long-range geometric properties of the structure are subject to a scaling law.

Consider a roughly spherical patch $\Omega$ of size $L$ in a disordered lattice. Let $S(\Omega)$ be the number of bonds in the patch and $J(\Omega)=\frac{\sum_{ij} J_{ij}}{S(\Omega)}$ be the average coupling within the patch. This raises the question of the behavior of these quantities at $L \rightarrow \infty$. Luck suggested that due to a scaling property, the new disorder may introduce a new index at large $L$, based on the estimates
\begin{equation}
\label{wandering}
(J(\Omega)- J_0) S(\Omega) \sim L^{d \beta},
\end{equation}
where $0 \leq \beta \leq 1$ is defined as the ``wandering" exponent capturing the effect of the geometric fluctuations, and $J_0$ is the expectation value of $J(\Omega)$ at infinite size. Then one gets the modification of the Harris criterion
\begin{equation}
\label{wandering-2}
\frac{J(\Omega)- J_0} {J_0} \sim L^{d (\beta-1)} \sim (x-x_c)^{d \nu (1-\beta)}
\end{equation}
as $S(\Omega)\sim L^d$. Thus, for disorder relevance, instead of $d \nu/2 < 1$ we have $d \nu (1-\beta)< 1$. The validity of the relevance criterion implied by Harris' arguments in general has been discussed in articles \cite{Zimani-1997,Aharony-1998,Herrmanns-1998}.

The structures conceptually related to these aperiodic models with a wandering exponent can be given by random lattices, such as the Poissonian Voronoi-Delaunay (PVD) triangulations \cite{Okabe-2000,Schliecker-2002} or the dynamically triangulated surfaces \cite{Ambjorn-1997}.
Although spin models have been extensively studied on quenched ensembles of triangulated lattices \cite{Espriu-1986,Janke-1993,Janke-1994,Janke-1995,Bialas-1999,Janke-2000,Janke-2000-2, Janke-2002},
the explicit effect of the lattice geometry,
namely the analysis of connectivity fluctuations and their influence on the wandering exponent,
was not presented until  \cite{Janke-2004, Vojta-2014,deOliveira-2008,Schrauth-2018}. In \cite{Janke-2004} the authors showed how the isotropic, power-law correlations of the connectivity fluctuations can change the wandering exponent formulated by Luck \cite{Luck-1993}.
Their results  show that PVD random lattices
have $\beta = 1/2$, thus leaving the Harris criterion unchanged.
On the other hand dynamical triangulations have $\beta \approx 0.72$
and modify the criterion according to Luck's arguments,
which in turn relaxes the relevance condition.  In \cite{Vojta-2014} also a random PVD lattice  was considered and fluctuations of connectivity were analyzed. The authors predicted the wandering exponent in arbitrary dimension $d$ and verified its value, $\beta=1/4$, 
for the Ising model on a  $d=2$ random PVD lattice. In this case the relevance condition is more restrictive  than Harris criterion.

In this article, we consider disordered networks described in the paper \cite{Sedrakyan-2017} and, tracing the reasoning of Harris and Luck, show that this disorder is always relevant and can lead to a change of the localization length exponent $\nu$. We numerically compute the behavior of the critical point fluctuations for quenched configurations of disordered networks and find that they are independent of the subsystem size $L$. In other words, instead of $\Delta x_c \sim L^{-d/2}$ as expected by the central limit theorem, we find that $\Delta x_c \sim L^{0}=const$. This indicates the value of the wandering exponent $\beta=1$ (see Eq.\ref{wandering-2}). 

\section{Harris-Luck criterion in the GKNS network
  model for the plateau transition in the IQHE.}
  
   According to the Harris criterion, the critical index in IQHE (Integer Quantum Hall Effect) is stable for any $\nu>1$. However, calculations show a change. A possible explanation for this phenomenon is the nature of the considered disorder. During the thermodynamic analysis, i.e. the study of the partition function, one has to calculate the average over the disorder. The averaging over configurations of disordered geometry together with the other disorders can be interpreted as quenched averaging over all geometry configurations. This affects the integration measure itself, rather than simply being an additional field to be integrated. Since the geometry of the system is generally the ``field" of gravity (like the metric tensor in general relativity), the above mentioned averaging is actually an introduction of quantum gravity. Moreover, in \cite{Conti-2021} it was shown that the notion of ``curvature" plays a key role in such systems. Since the integration measure for the gravitational field is not known, it is unclear how the central limit theorem should be reformulated for this particular case. The argument may seem very vague, but it can be checked directly.

Following Harris' procedure, we divide the system into subsystems of size $L$ with different geometry configurations and define critical points $x_{c,i}$ for each configuration $i$. If the critical point fluctuation $\Delta x_c^{(L)}$, induced by the new disorder for subsystem size equal to the correlation length $\xi \sim (x-x_c)^{-\nu}$, is stronger than $(x-x_c)$, then different subsystems are in different phases and the disorder is relevant.

The relation suggested by the central limit theorem is $\Delta x_c^{(L)} \propto L^{-d/2}$, which can be checked numerically. To compute $\Delta x_c^{(L)}$ for different values of $L$, we generate ensembles of different configurations for each of them and compute their critical points $x_{c,i}^{(L)}$. The $\Delta x_c^{(L)}$ is then simply the standard deviation of the $x_{c,i}^{(L)}$ distribution.

\begin{figure}
	\centering
	\subfigure[]{
            \includegraphics[width=0.22\textwidth]{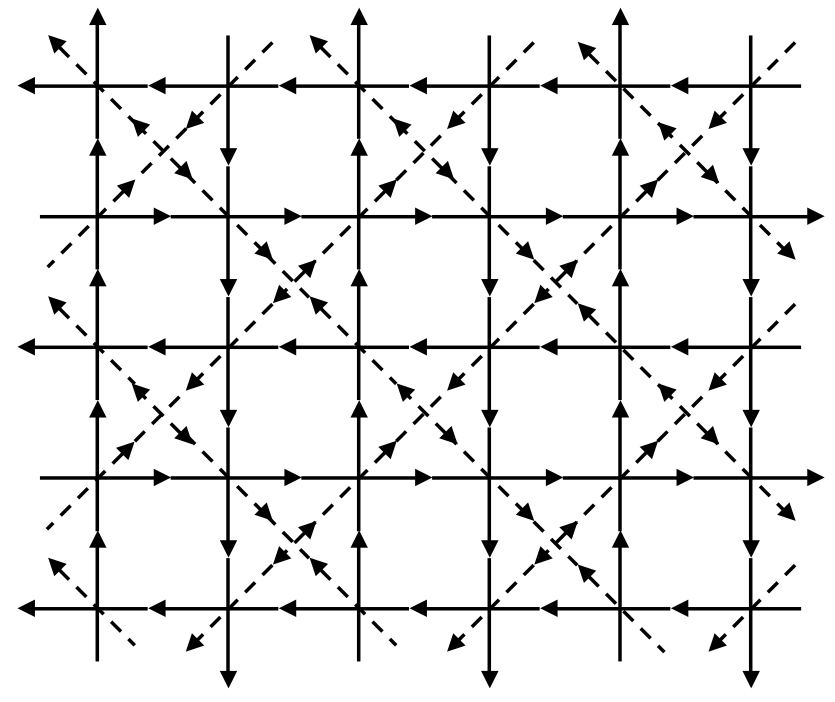} % ML12
            \label{fig:CC-network-a}
        } 
	\subfigure[]{
            \includegraphics[width=0.22\textwidth]{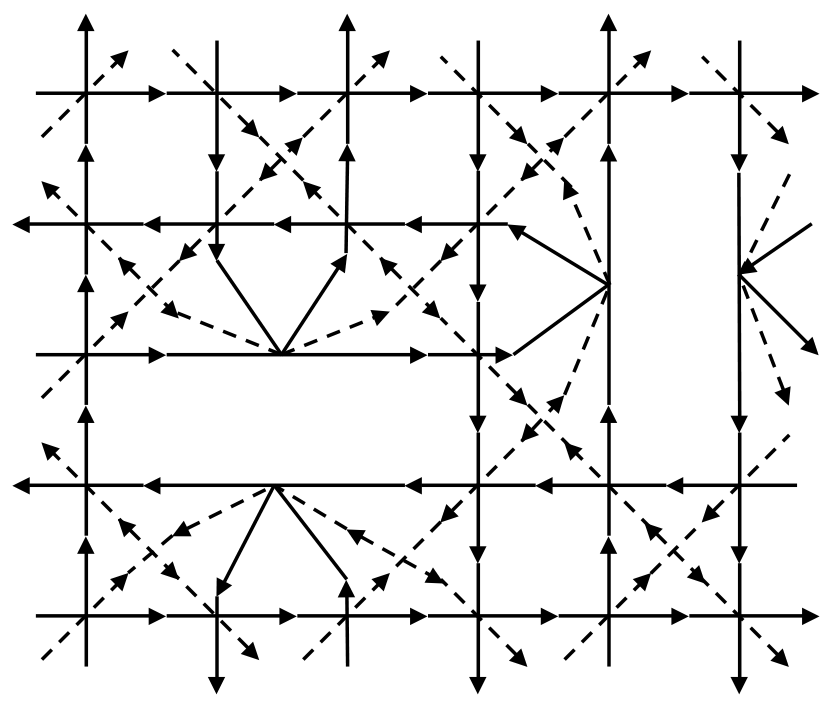} % random-lattice-31
            \label{fig:CC-network-b}
        } 
	\subfigure[]{
            \includegraphics[width=0.9\linewidth]{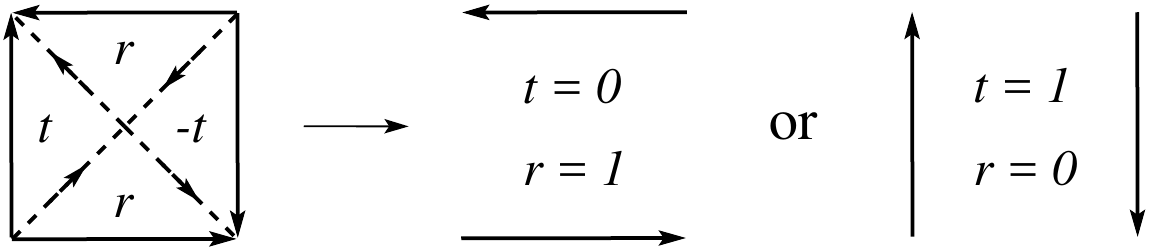}
            \label{fig:CC-network-c}
        }
	\caption{(a) Regular network of the CC model \cite{Chalker-1988}. 
    The scattering happens near the saddle points of the potential in \reff{CC-random}(a).
    The randomness of the potential is reflected only in the Aharonov-Bohm phase factors acquired by electrons in the hopping between neighboring saddle points.  
(b) Random network obtained from the regular network in (a) through the modifications shown in (c).  
(c) Schematic representation of the scattering replacement process in the CC model.}
	\label{fig:CC-network}
	\vspace{-.5cm} 
\end{figure}

\begin{figure}
%	\centering
	\subfigure[]{\includegraphics[width=0.23\textwidth]{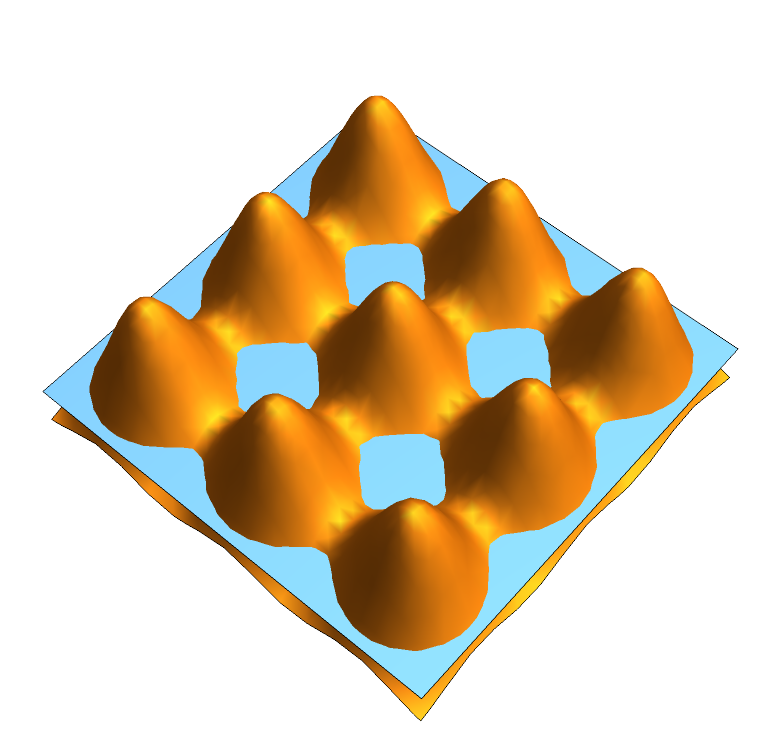}} 
	\subfigure[]{\includegraphics[width=0.23\textwidth]{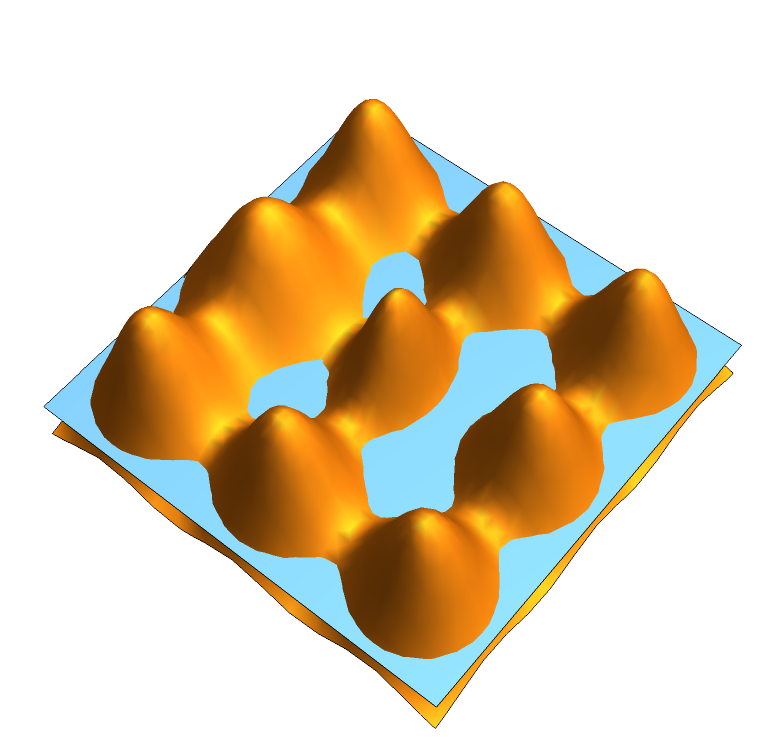}}
	\subfigure[]{\includegraphics[width=0.45\textwidth]{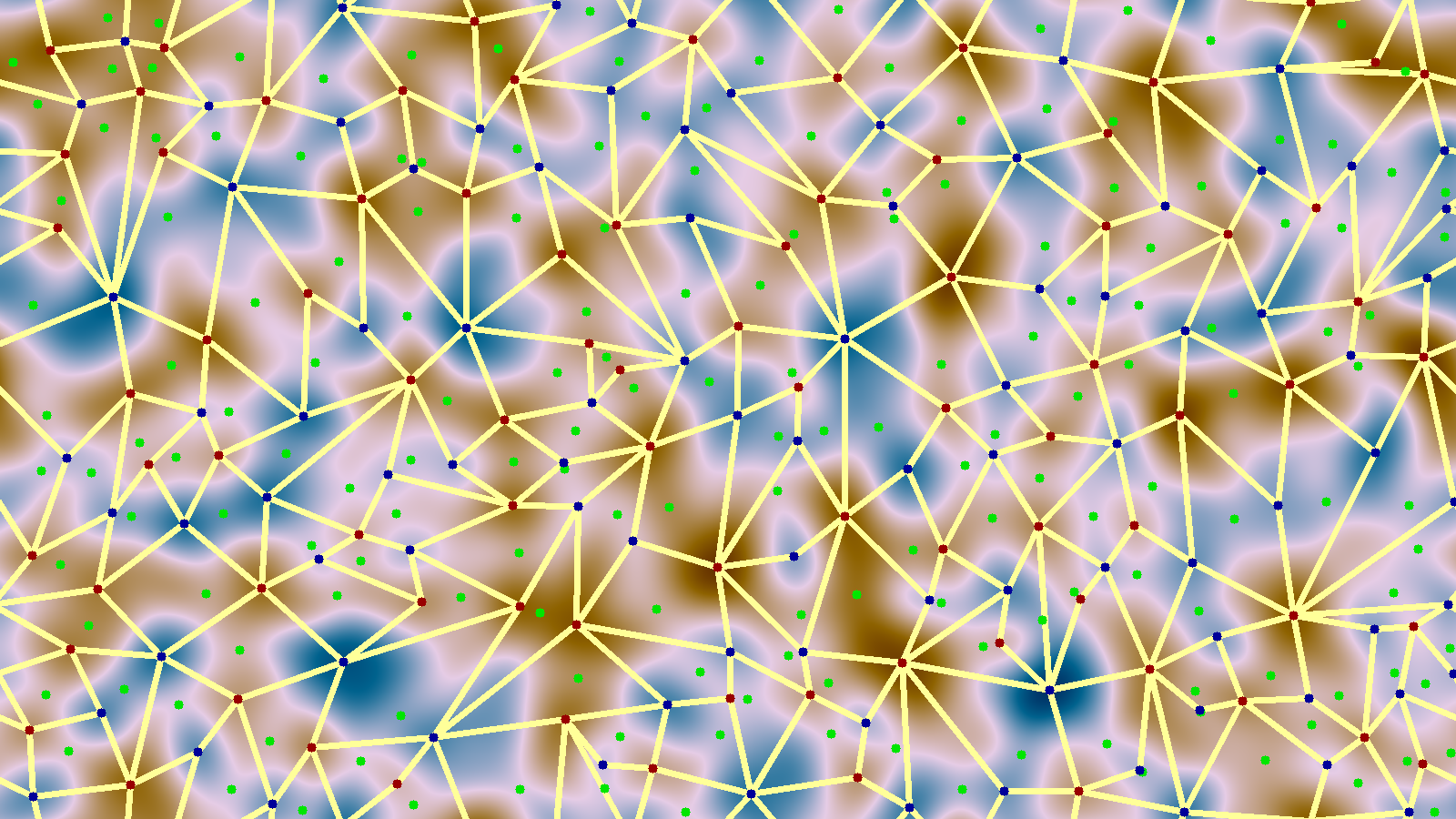}} 
	\caption{(a) Regular potential of the CC model \cite{Chalker-1988}. Scattering with parameters \( r \) and \( t \), as shown in \reff{CC-network-a}, occurs at the saddle points of the regular potential.  
(b) Modification of the potential in (a) by pressing a saddle point down into the Fermi sea. This corresponds to modifying the regular network in \reff{CC-network-b} using the operation shown in \reff{CC-network-c}.  
(c) Quadrangulated random surface (yellow lines) that appears on the disordered potential.
    The background represents the potential (blue to brown is to higher values)
    and the points are the extrema (red: maxima, blue: minima, green: saddle points.)}
%		after random operations of Fig.\ref{fig:R-marix}
	\label{fig:CC-random}
\end{figure}

\begin{figure}
%	\vspace{-.5cm}
	\centering
	\includegraphics[width=\linewidth]{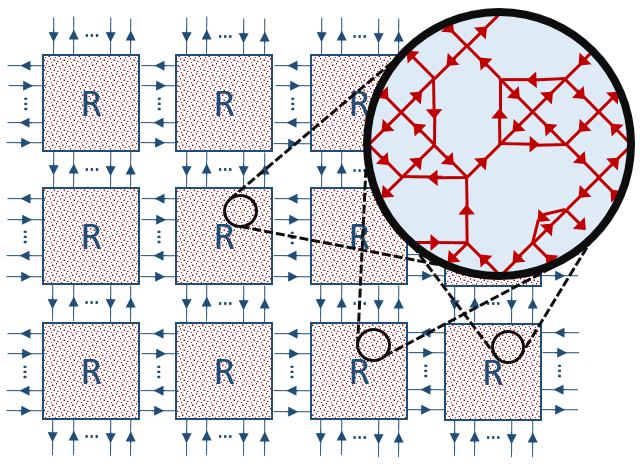}
	\caption{The structure of a system composed by
          replicating the same constituent subsystem block $R$, which is a disordered network with $p_n=1/3$.}
	\label{fig:realization}
	\vspace{-.7cm}
\end{figure}
The geometry disorder is formulated similarly to \cite{Sedrakyan-2017}. Namely,  we start from the regular CC network
model for the IQHE \cite{Chalker-1988}, which is a two-dimensional single-particle scattering network (Fig.\ref{fig:CC-network-a})
	with random phase shifts associated to the hoppings due to the magnetic field.
	Then normal scattering nodes given by reflection and translation amplitudes
$(r,t)$ are replaced by extreme scattering with $(r=1,t=0)$ or $(r=0,t=1)$, each with
probability $p_n$
\cite{Sedrakyan-2017,Sedrakyan-2019,Sedrakyan-2024} (see Fig.\ref{fig:CC-network-c}). This replacement generates a random network presented in Fig.\ref{fig:CC-network-b}.

The random networks described above can be associated with random potentials filled by a Fermi sea of particles up to the energy level \(E\) \cite{Conti-2021}. In this correspondence, the scattering nodes of the networks in \reff{CC-network}(a,b) (i.e., the network sites) correspond to the saddle points of the potential above the filled Fermi sea or lakes (see \reff{CC-random}(a,b)). The regular CC network in \reff{CC-network-a} is defined by the saddle points of the periodic potential shown in \reff{CC-random}(a).  

The modification operation in \reff{CC-network-c} is equivalent to ``pushing" saddle points down into the sea, as depicted in \reff{CC-random}(b), thus introducing fluctuations in the potential. These modifications generate random quadrangular surfaces (\reff{CC-random}(c)) formed by critical points of the random potential, which are equivalent to the scattering networks in \reff{CC-network-b}. The probability \( p_n \), defined for the GKNS networks above, is linked to the parameter \( p_c \in [0,0.5) \), which represents the ratio of saddle points submerged in the Fermi sea to the total number of saddle points in the random potential \cite{Conti-2021}.

The number of all possible configurations for a system of size $L$ is $3^{2L^2}$ (there are $3$ possibilities for each of the $2L^2$ initial scattering nodes: to stay ``as is" or to be replaced by $t=0$ or $r=0$ scatterings), which is too large to handle. Instead, we use the Monte Carlo approach and randomly pick configurations from the pool of all possibilities (\reff{realization}). This is done by creating a random network with $p_n=1/3$, which ensures a uniform distribution of picked samples throughout the pool.

To determine $x_{c,i}^{(L)}$ of a given configuration $i$ of size $L$, we must
first perform a quantum averaging over random phases. This can be done by
considering a system composed of the repeatedly tiled configuration $i$ with
different phase realizations, thus generating a network of size $k_v L \times
k_h L$ with $k_v, k_h \in \mathbb N$ (see Fig.\ref{fig:realization}). Then
$x_{c,i}^{(L)}$ is determined as the value $x$ for which the system has the largest correlation length or the smallest Lyapunov exponent $\Gamma_i^{(L)}(x)$.

In our calculations we have considered $18$ different systems of sizes $9 \leq
L \leq 100$. The ensemble for each $L$ consists of $20-40$ configurations $i$,
each tiled into a system of sizes $k_v L \approx 100$ and $k_h L \approx
5\cdot 10^6$. $x_{c,i}^{(L)}$ for each configuration $i$ is determined by
analyzing the Lyapunov exponents $\Gamma_i^{(L)}(x)$ at $\sim 30$ values of
$x$ near $x_c$. Each $\Gamma_i^{(L)}(x)$ is computed as a patch average
over
$\sim 500$ calculations of $\Gamma_i^{(L)}(x)_j$
with same geometry (index $i$), but different values of the
  random phases (index $j$). Then the
variance $\Delta x_c^{(L)}$ can be determined for each system
size.

The uncertainty analysis is a crucial point in the problem. First, each $x_{c,
  i}$ is determined as the minimum point of $\Gamma^{(L)}_i(x)$, which is
ideally defined on an infinite system. In our case we have patches of measured
values $\widetilde{\Gamma}^{(L)}_i(x)_j$. Then $\Gamma^{(L)}_i(x) =
\widetilde{\Gamma}^{(L)}_i(x) \pm \delta\Gamma^{(L)}_i(x)$ with
\begin{equation}
\begin{split}
	\widetilde{\Gamma}^{(L)}_i(x) &= \inb<{ \widetilde{\Gamma}^{(L)}_i(x)_j}>_j  \quad,\\ 
	\delta\Gamma^{(L)}_i(x) &= \sqrt{\sum_j\inb({ \widetilde{\Gamma}^{(L)}_i(x)_j - \widetilde{\Gamma}^{(L)}_i(x)})^2} \bigg/ \# j \quad,
\end{split}
\end{equation}
where $\inb<\cdot>_j$ means averaging over $j$, and $\# j$ is the patch size.

This then propagates into the uncertainty of $x_{c, i}^{(L)}=\tilde{x}_{c,
  i}^{(L)}\pm\delta x_{c, i}^{(L)}$, which is obtained by a fitting procedure.
It becomes essential when calculating $\Delta x_c^{(L)}$,
\begin{equation}
\begin{split}
	\Delta x_c^{(L)} &= \sqrt{\sum_i\inb({x_{c, i}^{(L)} - \inb<{x_{c,i}^{(L)}}>_i})^2} \bigg/ \#i = \\
	&= \sqrt{\inb.{\widetilde{\Delta x}_c^{(L)}}.^2 - 2D_L^2} 
	\pm \frac{\sqrt{2}}{2} D_L \quad,
	\label{eq:uncert_form}
\end{split}
\end{equation}
where
\begin{equation}
\begin{split}
	\widetilde{\Delta x}_c^{(L)} &= \sqrt{\sum_i\inb({\tilde{x}_{c, i}^{(L)} - \inb<{x_{c, i}^{(L)}}>_i})^2} \bigg/ \#i \quad, \\ 
	 D_L&=\sqrt{\sum_i {\delta x_{c,i}^{(L)}}^2} \bigg/ \#i \quad.
	\label{eq:uncert_exprs}
\end{split}
\end{equation}
The calculations are done using a first order approximation on $\delta
x_{c,i}^{(L)}$. Here we took advantage of the fact that the value of
$\inb<{x_{c, i}^{(L)}}>_i$ is known in advance, otherwise we would have
$\inb<{x_{c, i}^{(L)}}>_i = \inb<{\tilde{x}_{c, i}^{(L)}}>_i \pm D_L$, with
further propagation to \refe{uncert_form}. A
noteworthy point is that the uncertainties induce a shift of $\Delta x_c^{(L)}$, which should be taken into account.

\begin{figure}
    \centering
    \includegraphics[width=\linewidth]{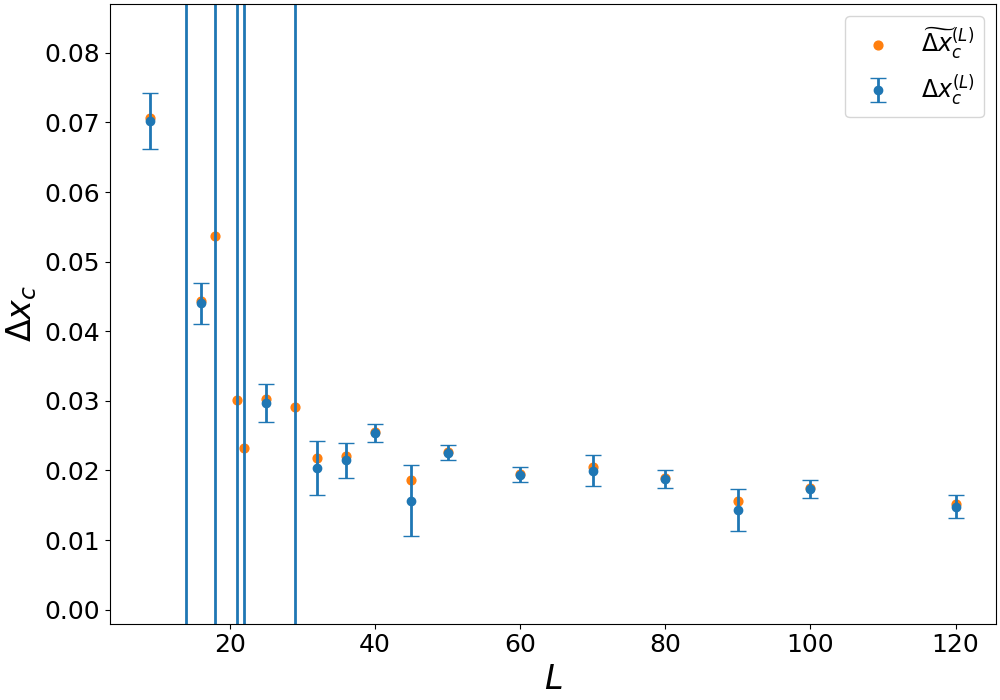}
    \caption{Dependence of $\widetilde{\Delta x}_c$ and $\Delta x_c$ on $L$. 		Uncertainties propagate from the minimum detection of $\Gamma$ curves. 		The $\Delta x_c$ are undefined if $D_L\sim \widetilde{\Delta x}_c^{(L)}$.}
    \label{fig:harris_data}
    \vspace{-.5cm}
\end{figure}

The results obtained for $\Delta x_c^{(L)}$ are shown in
\reff{harris_data}. As can be seen, $\Delta x_c$ decreases with increase of
$L$ reaching a non-zero plateau, whereas according to the central limit theorem it
should decrease to $0$ in a power law (namely as $L^{-1}$ in our case). Thus,
close to criticality, the critical point fluctuation of correlation length
sized subsystems is $\Delta x_c \sim (x-x_c)^0 = 1 \gg x-x_c$. This
indicates that the introduced disorder of the network is relevant and can lead
to a change of the CC model value $\nu \simeq 2.56$.

\section{Comparison to other models with disorders of connectivity.}

In the papers \cite{Janke-2004,Vojta-2014} the random geometry of lattices was studied numerically based on their connectivity. In \cite{Janke-2004} it was found that the wandering
exponent for the case of PVD triangulation is $\beta =1/2$, so the Harris
criterion remains valid, while in \cite{Vojta-2014} $\beta =1/4$ and the modified Harris criterion for relevance gets more restrictive. On the other hand, for the connectivity fluctuations
based on dynamic triangulation, $\beta \simeq 3/4$ was obtained in \cite{Janke-2004}, indicating
that disorder is relevant for all spin chains and that the correlation length
index $\nu$ can be changed. It seems that it is not only the wandering exponents that distinguish the two types of random lattices, but also their distributions $P(q)$ of connectivity $q$. In \cite{Janke-2004} it was confirmed that in the PVD triangulation model $P(q)$ is Poissonian, while the dynamically triangulated lattice has an exponential distribution at large $q$. It is consistent with earlier statements that at large coordination numbers the distribution $P(q)$ decays as $\exp(-\sigma q \ln q)$ with $\sigma \simeq 2$ for PVD random lattices \cite{Itzikson-1984}, while for dynamical triangulations it decays slower, proportional to $\exp(-\sigma q)$ with $\sigma = \ln (4/3) \simeq 0. 3$ \cite{Ambjorn-1997}.

\begin{figure}
	\centering
	\subfigure[]{\includegraphics[width=0.45\textwidth]{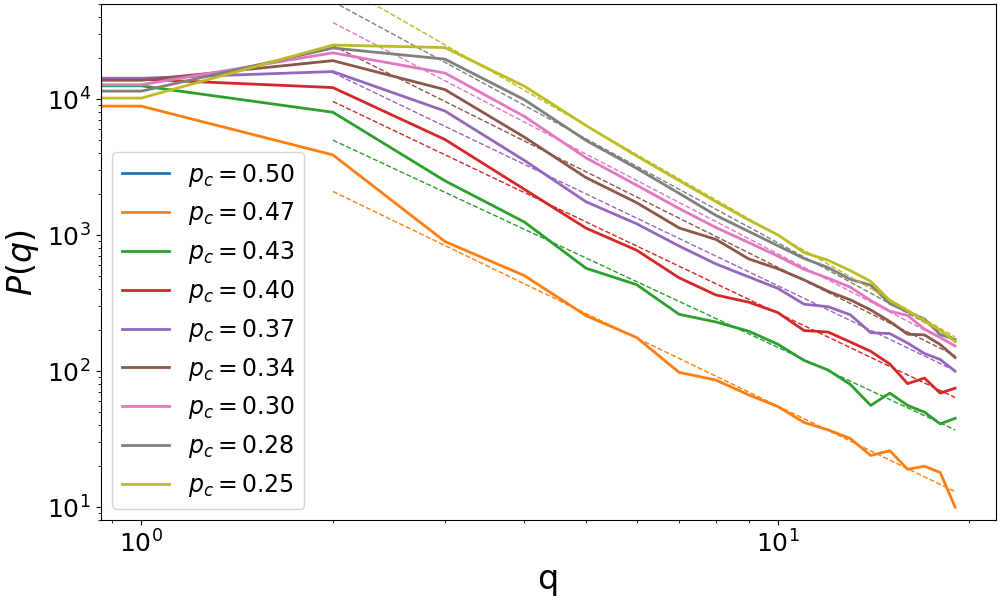}} 
	\subfigure[]{\includegraphics[width=0.45\textwidth]{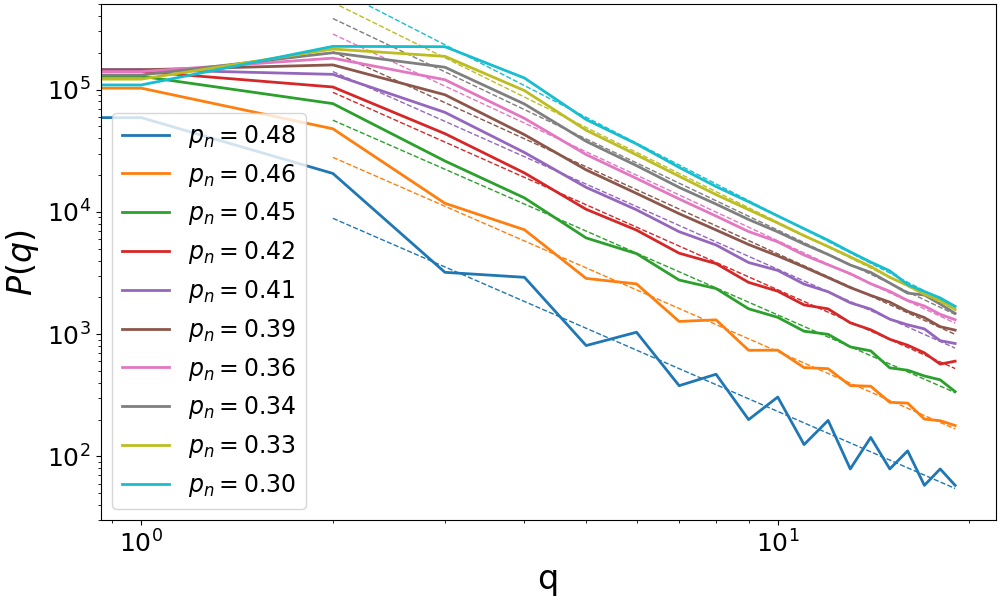}} 
	\caption{(a) Random-potential-induced scattering networks for various values of $p_c$. (b)Distributions $P(q)$ of network node connectivity $q$ (log-log plots) for different parameterizations (solid lines), and the corresponding power law fits (dashed lines). }
	\label{fig:distr}
\end{figure}

In our model \cite{Sedrakyan-2017}, the network can be mapped onto a random quadrangular surface \cite{Conti-2021} with variable connectivity $q$, which has a distribution $P(q)$. Our current analysis of the distributions $P(q)$ \cite{Conti-2021} shows that $P(q)$ decays in algebraic manner, $P(q) \simeq q^{-\kappa}$ (\reff{distr}(a)). For the disordered network model \cite{Sedrakyan-2017} $\kappa=2.5$ at $p_n=1/3$ (\reff{distr}(b)). For both the disordered networks (\reff{distr}(b)) and the random-potential-induced scattering networks (\reff{distr}(a)) \cite{Conti-2021}, all the presented distributions $P(q)$ are also well approximated by the power law regardless of the parameters $p_n$ and $p_c$, with $\kappa$ in the range $2.2 \sim 2.7$ depending on the value of $p_{n/c}$.
Hence we have a random network model that is essentially different from PVD and dynamically triangulated lattices.
Complex networks with a similar property of a power law distribution of connectivity,
were studied previously (see \cite{Albert-2002} for a review)
and they describe a wide range of problems in nature and society.
The models with a power law distribution of connectivity are called ``scale free" \cite{Barabas-1999}.
This is consistent with the previously obtained result $\Delta x_c \propto L^0$.
There is a large number of such networks, including the World-Wide Web \cite{Albert-1999},
the Internet \cite{Faloutsos-1999}, or metabolic networks \cite{Jeong-2000}.
Our model \cite{Sedrakyan-2017} augments the set of scale free network models.
Note that, unlike PVD and dynamical triangulations, our network is quadrangulated (\reff{CC-random}(b)).

\section{Summary.} 
We demonstrated that the usual derivation of Harris criterion in case of
geometric disorder does not hold. It happens because disorder of geometry as
in \cite{Sedrakyan-2017} results in deviations of critical point positions, which is not
explained by the classical formulation of the central limit theorem. This
means that this type of geometric disorder can be relevant and can lead to a change of the
critical indices of the GKNS system.

The power law asymptotics of the
distribution $P(q)$ of connectivities essentially distinguishes GKNS network
from other models of random lattices, namely PVD and dynamical
triangulations. It indicates that GKNS network belongs to a different
universality class than these two models,
namely to the class of scale free complex networks \cite{Barabas-1999}.
The fact that models on dynamically
triangulated lattices change their critical indices was known decades ago from
studies of non-critical string theories
\cite{Migdal-1986,Kazakov-1988,Kazakov-1988-2,Kazakov-1989,Polyakov-1988}. Now
we have observed the mechanism of changing the localization length index in
GKNS disordered networks.  Different distributions of connectivities
in the case of dynamical triangulations and GKNS networks mean that these lattices
have different measures in the disorder averaging process. This may have
interesting consequences for non-critical string theories. The mentioned
differences also explain the discrepancy between the values of the numerically
calculated localization length indices in IQHE plateau transitions on the PVD
random lattice \cite{Bera-2024} and a GKNS network \cite{Sedrakyan-2017}.
Papers \cite{Janke-2004,Vojta-2014} report $\beta=1/2$ and $\beta=1/4$,
implying that the Harris criterion for relevance is not satisfied
for the CC model ($\nu \approx 2.6$) on a PVD random lattice. As a result
the localization index should remain the same as in the CC model,
which was observed in \cite{Bera-2024}.
But it can be changed in the GKNS network
\cite{Sedrakyan-2017}, as $\beta=1$.

\section*{Acknowledgments.} 
The authors thank F. Evers, I. Gruzberg, and
T. Hakobyan for helpful discussions. The research was supported by Armenian
HESC Grants Nos. 21AG-1C024 (AS), 24RL-1C024 (HT) and
24FP-1F039 (HT, AS). A.K. acknowledges
funding from Deutsche Forschungsgemeinschaft (DFG) via Research Unit 2316. The
authors also acknowledge the Paderborn Center for Parallel Computing, where
all numerical calculations were performed.

\bibliography{refs}
 
\end{document}